\documentclass[prb,preprint,samsmath,amssymb,floatfix,superscriptaddress,nofootinbib,aps]{revtex4}
\usepackage{graphicx}
\usepackage{epstopdf}
\usepackage{amsmath}

\usepackage{color}

\begin{document}

\title{{\it Predicting} Patchy Particle Crystals:\\ Variable Box Shape
  Simulations and Evolutionary Algorithms}

\author{Emanuela Bianchi} 
\affiliation{Institut f\"ur Theoretische Physik and Center for
  Computational Materials Science (CMS), Technische Universit\"at
  Wien, Wiedner Hauptstra{\ss}e 8-10, A-1040 Vienna, Austria}
\author{G\"unther Doppelbauer}
\affiliation{Institut f\"ur Theoretische Physik and Center for
  Computational Materials Science (CMS), Technische Universit\"at
  Wien, Wiedner Hauptstra{\ss}e 8-10, A-1040 Vienna, Austria}
\author{Laura Filion} 
\affiliation{Soft Condensed Matter, Debye Institute for NanoMaterials
  Science, Utrecht University, Princetonplein 5, 3584 CC Utrecht, The
  Netherlands and Department of Chemistry, University of Cambridge,
  Lensfield Road, CB2 1EW, United Kingdom}
\author{Marjolein Dijkstra} 
\affiliation{Soft Condensed Matter, Debye Institute for NanoMaterials
  Science, Utrecht University, Princetonplein 5, 3584 CC Utrecht, The
  Netherlands}
\author{Gerhard Kahl} 
\affiliation{Institut f\"ur~Theoretische Physik and Center for
  Computational Materials Science (CMS), Technische Universit\"at
  Wien, Wiedner Hauptstra{\ss}e 8-10, A-1040 Vienna, Austria}

\begin{abstract}
We consider several patchy particle models that have been proposed in literature and we 
investigate their candidate crystal structures in a systematic way. 
We compare two different algorithms for predicting crystal structures: 
(i) an approach based on Monte Carlo simulations in the isobaric-isothermal 
ensemble and (ii) an optimization technique based on ideas of evolutionary algorithms. 
We show that the two methods are equally successful and provide consistent results 
on crystalline phases of patchy particle systems.
\end{abstract}


\maketitle

\section{Introduction}
\label{sec:intro}

Over the last number of years considerable effort has been dedicated to 
{\it predict} the crystalline phases for a wide variety of model systems.
In the case of strongly interacting systems, such as atomic and
molecular ones, much of the phase behavior is governed by the zero
temperature case. In such situations, techniques which minimize the
thermodynamic potential (e.g., genetic algorithms or Monte Carlo basin
hopping simulations) have proven to be very useful in predicting the
ground state structures~\cite{Wales99,WoodleyC08,OganovG06}. Motivated by
these successful approaches, the aforementioned optimization procedures
have been extended to soft matter systems and turned out to be
efficient and robust techniques, suitable for a reliable prediction of
crystalline structures at zero temperature for a broad variety of
systems~\cite{GottwaldKL05,GA-shoulder08,Kah09,Gernot2009,Kahn2010}.
Optimization strategies search for particle arrangements that minimize the thermodynamic potential of a system at zero temperature
and identify them as candidate equilibrium structures.  
For finite temperatures this minimization criterion cannot be completely trusted as one cannot safetly neglect the entropic contributions to the thermodynamic potential. Indeed, 
it has been shown that, even though the ground state structures provide a good guess for the finite temperature candidate structures, crystal phases that are local minima of the thermodynamic potential
can be thermodynamically stable at finite temperature~\cite{Doppelbauer-LMC2011,Doppelbauer2011}.
Optimization techniques which combine quasi-Newton local and global optimization steps
can be successfully applied to soft matter systems thanks to the smoothness of the inter-particle interaction, 
which is needed to guarantee that the derivatives involved in the minimization 
procedures of the potential energy are continuous.

In the case of hard (colloidal) particles, the situation is
considerably different and much more difficult. The phase behavior 
in purely hard systems is completely governed by the entropic 
contribution to the free energy.
Hence, when applying minimization techniques to hard systems
in order to identify the stable ordered equilibrium structures, 
the question regarding what to minimize arises.
Frequently in the past, the maximum packing fraction criterion has been 
used as it minimizes the Gibbs free energy at infinite pressure~\cite{Torquato,Filion-ga,Joost-prl}. 
For finite pressures, the entropic term cannot be neglected and hence this criterion is not fully reliable.
For instance, for binary hard-sphere mixtures,  crystalline structures which are not the best packed ones 
exist as stable phases in the phase diagram~\cite{Filion-ga}.

Recently, a method has been proposed to cope with
hard-particle systems~\cite{Filion-npt}. This approach is a statistical sampling method 
used as a search strategy for candidate crystals structures of given systems.  It is
based on simple Monte Carlo
simulations of only a small number of particles: simulations are
carried out at constant temperature and pressure in a simulation box, 
whose shape is free to fluctuate~\cite{Filion-npt,Yip_83}. In
Ref.~\cite{Filion-npt}, Filion {\it et al.}  demonstrate that the
method successfully predicts both the infinite pressure as well
as the finite pressure crystal phases for a variety of hard-core
systems including binary hard-sphere mixtures, oblate hard
spherocylinders, hard asymmetric dumbbells and hard bowl-shape
particles. Additionally, the method turned out to be successful for
systems where particles interact via long range Coulomb interactions
and Lennard-Jones interactions. 
Hence, the variable box shape approach also offers the
possibility of predicting crystal phases at finite temperatures, as
opposed to the zero temperature minimization techniques. However,  
we emphasize that these techniques determine only
candidate crystal phases: subsequent full free energy calculations are
still required to identify the stable phases and to draw the complete
phase diagram~\cite{Marechal-2010,Marechal-2011,Doppelbauer-LMC2011,Doppelbauer2011}.

The present paper is dedicated to the prediction of crystal structures of patchy
particle systems. To be more specific, we study a variety of patchy
models that have been proposed in literature~\cite{Patchy_rev2011} and
search for their crystal structures, using both the variable box shape
simulation method as well as an optimization approach based on ideas
of evolutionary algorithms. Our motivation resides in the fact that patchy
particles have become a class of promising colloidal particles that
are able to self-assemble as building units of future
materials~\cite{Glotz_Solomon_natmat,zhangglotzer,Glotz_04}, with a
host of wide-spread applications, ranging from photonic
crystals to biomaterials. Thus, controlled synthesis and abundant production
of colloidal particles carrying a specific
(chemical or physical) pattern on their surfaces has become a hot
topic in soft matter
physics~\cite{Mohwald2006,Cho2008,Kraft2009,Kretz2009,VelevKretz,Granik-Kagome,Patchy_revExp}. 
A {\it reliable} prediction of the ordered equilibrium phases of such
systems represents therefore an important element in designing larger,
functional units with desired properties.

While, for patchy particles, an evolutionary algorithm approach has recently 
demonstrated its power in successfully predicting candidate crystal
phases that have shown to be stable at finite temperatures~\cite{JPCM10,Doppelbauer2011}, 
in this contribution we apply for the first time the variable box shape technique to model systems with
 strong directional interactions. Hence, in the body of the paper we
discuss candidate crystals obtained via variable box shape simulations
for different models of patchy systems. Our results are compared 
--whenever possible -- with the available literature, in an effort to validate
this  method. For some of the investigated models, we directly compare results 
from the variable box shape simulation technique to those from the evolutionary algorithm method.
We show that the two methods provide consistent results for the investigated patchy models.

The paper is organized as follows. In section~\ref{sec:models}, we
introduce the selected models for patchy particles. In
section~\ref{sec:methods}, we briefly describe the numerical methods,
referring for more details to previous
publications. Section~\ref{sec:results} is dedicated to the discussion
of the ordered crystal structures for a selection of different patchy
particle models. Finally, in section~\ref{sec:conclusions}, we draw
our conclusions and provide a comparison between the two algorithms
when applied to predicting ordered phases of patchy systems.

\section{Models}\label{sec:models}

We consider several patchy particle models that
have been proposed in literature during the past years. We distinguish
two main categories: discontinuous, square-well type potentials and
continuous, Lennard-Jones type interactions. For all the investigated
models we focus on the single bond per patch regime.

\subsection{Orientational Square-Well Models}

Most of the patchy particle models proposed and used in literature are
based on hard-core particles whose surfaces are decorated by a fixed
number of bonding patches interacting via a square-well potential.
Here, we consider two orientational square-well models: the
``sticky spots'' model~\cite{Bianchi06} and the Kern-Frenkel
model~\cite{KernModel03}.

The sticky spots model consists of hard-spheres carrying a small
number of attractive points arranged in a regular geometry on the particle surface. 
The pair potential
between two particles is given by the sum of an isotropic hard-core
repulsion of diameter $\sigma$ and a site-site attraction. 
Sites on different particles interact via a square-well
potential of depth $\epsilon$ and attraction range
$\delta=0.119\sigma$; this choice of $\delta$ guarantees that each
site is involved at most in one bond~\cite{Bianchi06}.  In Section
\ref{sec:results}, we show results for the case of six sticky sites per
particle, thus each particle can form up to six
bonds. Consequently, the average energy per particle, $e$, can vary
from $0$ (system of monomers) down to $-3 \epsilon$ (fully bonded
system).

In the Kern-Frenkel model, pairs of particles interact via a
square-well potential of depth $\epsilon$ and attraction range
$\delta$, modulated by a function that depends only on the relative
orientation of the attractive patches located on the two interacting
particles. This function is zero when the patches do not feel each
other; in this case particles only experience the
repulsion due to the hard-core of diameter $\sigma$. Otherwise, when
the patches can feel each other,
the modulating function is equal to unity. Two patches on different
particles feel each other when (i) particles are separated by a
distance smaller than $\sigma+\delta$, and (ii) the vectors connecting the 
center of the particle with the center of the patch form an angle less than a maximum angle, $\theta_{\rm max}$,
with the vector connecting the centers of the two particles. 
By appropriately choosing $\theta_{\rm max}$ and $\delta$, 
multiple bonds between patches can be avoided and the single bonding regime 
is guaranteed~\cite{Romano10}.  In Section \ref{sec:results}, we show results
for the case of particles with four patches arranged in a
tetrahedral geometry. The patch-patch attraction range and the patch angular extension 
are chosen to be $\delta=0.24\sigma$ and $\theta_{\rm  max}={\rm acos}(0.92)$,
respectively. The phase behavior of this particular
system has been investigated in Refs.~\cite{Romano10,Romano11}.

\subsection{Orientational Lennard-Jones Models}

More realistic models for patchy particles describe the directional
pair interaction via continuous and smooth pair potentials, which
are in general longer ranged than their orientational, square-well
counterparts. We consider a patchy particle
model first introduced in Ref.~\cite{Doye07cryst}. 
In the model, the repulsion between two particles is given by an isotropic
Lennard-Jones repulsive core, while the directional patch-patch
attraction is specified by a Lennard-Jones attraction of depth $\epsilon$,
modulated via a Gaussian-shaped angular decay.
Provided that the patches are sufficiently narrow, the single bonding
condition is guaranteed. We choose the following interaction
parameters: the cut off of the
attractive tail is $r_c=2.5\sigma$ and the width of the Gaussian
modulation is $\omega=0.3$ rad~\cite{Noya07,Noya2010}. In Section
\ref{sec:results}, we show results for the cases of four and six
patches per particle arranged in a tetrahedral and an octahedral
geometry, respectively. The phase diagrams for such systems have
been investigated in Refs.~\cite{Noya07,Noya2010}.

Other geometrical arrangements of the patches within the orientational
Lennard-Jones pair potential can be introduced via an additional
geometrical parameter $g$, which controls the patch positions on the
particle surface~\cite{Doppelbauer2011}. Here we focus on the case of 
four patches per particle, with $r_c=1.9\sigma$ and $\omega=0.22$ rad. 
The parameter $g$ is the central angle between a patch chosen as the pole of 
the particle and any of the other three patches.
Hence, $g \approx 109^{\circ}$ specifies a regular tetrahedral arrangement.
By varying $g$, the arrangement of the four attractive patches ranges from a rather
compressed to a rather elongated tetrahedron, so that it is possible
to study the effect of the patch geometry on the crystal structures.

\section{Methods}\label{sec:methods}

To predict crystalline structures, we use two different methods: the variable box shape simulation method of Ref.~\cite{Filion-npt} and an optimization approach based on ideas of evolutionary algorithms~\cite{GottwaldKL05,JPCM10,Doppelbauer2011}. 

The identification of the various ordered structures and the comparison of the results obtained from the two different algorithms are based both on visual inspection and on a bond-order parameter analysis~\cite{Steinhardt,Dellago}.

\subsection{Variable Box Shape Simulations}

The algorithm is based on Monte Carlo (MC) simulations in the isothermal-isobaric ensemble: the initial state point  is chosen to be in the fluid phase and then, at constant temperature $T$,  the pressure $P$ is increased step by step  until the system converges to a final crystalline structure. We treat the simulation box as a unit cell and we allow its shape to fluctuate in order to avoid any bias of the crystal structure.  As a consequence of the chosen box sizes, we work with extremely small numbers of particles $N$. In this paper, the particle numbers in the simulation box range typically from 4 to 8, but we also run simulations with up to 12 particles in the unit cell.  
Each MC step consists on average of $N$ trial particle moves and one trial volume change, where the corresponding acceptance rules are given by the Metropolis algorithm.  
A particle move is defined as both a displacement in each direction of a random quantity distributed uniformly between $\pm \delta r$ and a rotation around a random axis of a random angle distributed uniformly between $\pm \delta \theta$. A volume move is given by a trial change of a randomly chosen component of a randomly chosen vector of the simulation box by a random quantity uniformly distributed between  $\pm \delta v$.  The chosen values for the trial changes are $\delta r =   0.05 \sigma$,  $\delta \theta = 2\delta r/\sigma$ rad, and $\delta v = 0.02\sigma$, but they are allowed to change during the simulation runs in order to keep the acceptance rates of both types of trial moves between $30\%$ and $40\%$.  The size of the trial moves is changed according to the following rule: if the acceptance probability --calculated every $10^5$ Monte Carlo steps-- of a particle (or volume) move is smaller than 30\% then $\delta r=0.95 \delta r$ (or $\delta v=0.95 \delta v$); if the acceptance rate is bigger than 40\% then $\delta r=1.05 \delta r$ (or  $\delta v=1.05 \delta v$). Upper and lower limits for the step size are fixed in order to prevent extreme fluctuations in the fluid regime, namely $\delta r_{\rm min}$ (or $\delta v_{\rm min}$) = 0.001 and $\delta r_{\rm max}$ (or $\delta v_{\rm max}$) =0.5. Typically, both $\delta r$ and $\delta v$ increase abruptly in the fluid phase, because the initial configuration is in a very dilute regime; on progressively increasing pressure both $\delta r$ and $\delta v$ equilibrate fast to a value around $0.05\pm 0.05\sigma$ (bigger step sizes for higher temperatures).

During the runs, we prevent the box from undergoing extreme distortions by using a lattice reduction technique~\cite{GottwaldKL05} in order to avoid extremely time consuming energy calculations. Moreover,  we impose a lower bound on the length of all the lattice vectors: each of the lattice vectors must be longer than $1.5\sigma$. In this way, we avoid extreme elongations of the simulation box, in which the particles tend to form columns so that they only interact with the periodic images in one of the lattice directions.

Once the number of particles, the temperature and the initial pressure are chosen, we run  several MC simulations in parallel, starting from different initial conditions. The pressure is increased step by step, using on average 100 pressure steps from the initial to the final pressure; for each pressure value we perform $10^6$-$10^7$ MC steps.  We distinguish two ranges for the final pressure: (i)  low pressure values, ranging from  0.01 to 10 (in units of $\epsilon/\sigma^3$) and (ii) high/intermediate pressure values, ranging from 10 to 200. Different temperatures in the range from 0.10 to 0.20 (in units of  $\epsilon/\kappa_B$) are considered. For each state point,  we check if convergence to a certain final structure occurs over the last part (about one third) of each MC run.

We note here that this method assumes that states which are stable for large systems are at least metastable for small systems, a point we feel to be largely validated by the fact that the method works well for the large variety of systems it has been tested for~\cite{Filion-npt}.  We also note that the small system sizes aid us in exploring phase space in two ways: (i) the fluctuations in density at a fixed pressure are larger than for larger systems; as such, near coexistence, the system frequently crosses the fluid-solid phase boundary and has a high probability of finding the stable phase, (ii) the small systems allow for large rearrangements of particles, and hence significant changes in the crystal structure which would not be possible for large systems.

\subsection{Evolutionary Algorithm}

In an effort to identify the ordered equilibrium structures formed by
the patchy particles at vanishing temperature we use
optimization techniques that are based on ideas of evolutionary
algorithms~\cite{GottwaldKL05} (EA). Working at fixed particle number (per unit
cell) and fixed pressure $P$, the Gibbs free energy $G$ is optimized
with respect to (i) the lattice vectors specifying the unit cell, (ii)
the positions of the particles within the unit cell, and (iii) the
orientations of these particles. At zero temperature, $G$ is reduced to the
enthalpy: $G \rightarrow H = U + PV$, $U$ being the lattice sum and
$V$ the volume of the system. In this contribution we use a
phenotype implementation~\cite{Dea95} of such an algorithm, combining
global optimization steps with local ones, as specified in Ref.~\cite{JPCM10} for the two-dimensional case.
For technical details about the generalization to the three-dimensional case,
we refer the reader to Ref.~\cite{Doppelbauer2011}. In
the optimization runs, up to 8 particles per unit cell are
considered. A population of usually ten individuals (each of them
corresponding to an ordered candidate structure) is iterated along an
evolutionary path via the usual mating, mutation and local minimization operations
performed on the individuals (for details cf Ref.~\cite{JPCM10}). Throughout the
optimization runs, the parameters of all these individuals are
recorded. Among those, the one with the lowest value for the Gibbs
free energy is considered as the final solution (global minimum) for
this particular run; in addition, further structurally different~\cite{Steinhardt}
 local minima on the $G$-landscape are recorded. At least
three and up to ten independent optimization runs are carried out in
parallel for a given state point in order to ensure consistency.

\section{Results}\label{sec:results}

In the following, we discuss candidate crystals obtained via the variable box shape simulation approach
for all the patchy models described in Sec.~\ref{sec:models}. Whenever possible, we compare our results with 
the available literature, in an effort to validate the method for systems characterized by strong directional interactions. 
Moreover, for the orientational Lennard-Jones models, we directly compare results from the variable box shape simulation technique 
to those from the evolutionary algorithm method. All the lattice structures shown in the figures are MC output data, 
while for the visual representation of the EA output data we refer to Refs.~\cite{Doppelbauer2011,Doppelbauer-LMC2011}.
The comparison between results from the two methods is reported in Table~\ref{table:tab-ga} and~ Fig.\ref{fig:tetra}.

\subsection{Orientational Square-Well Models}
We first consider a particular realization of the Kern-Frenkel model with four patches,
whose phase diagram has been extensively studied in Refs.~\cite{Romano10,Romano11}. 
In these papers, the following stable phases have been studied: the Face-Centered-Cubic (FCC) structure 
at high densities, the Body-Center-Cubic (BCC) crystal at intermediate densities, and two, tetrahedrally arranged, 
open structures, i.e. the Diamond-Cubic (DC) and the Diamond-Hexagonal (DH) 
crystals~\cite{Romano10,Romano11}.
In our MC simulations, we observe almost all the previously predicted crystal phases, only instead of the BCC lattice 
we identify a Body-Centered-Tetragonal (BCT) phase. In addition, we observe two Hexagonal-Close-Packed 
(HCP) crystals with different bonding patterns. 

\begin{figure}[h]
\includegraphics[width=10cm, clip=true]{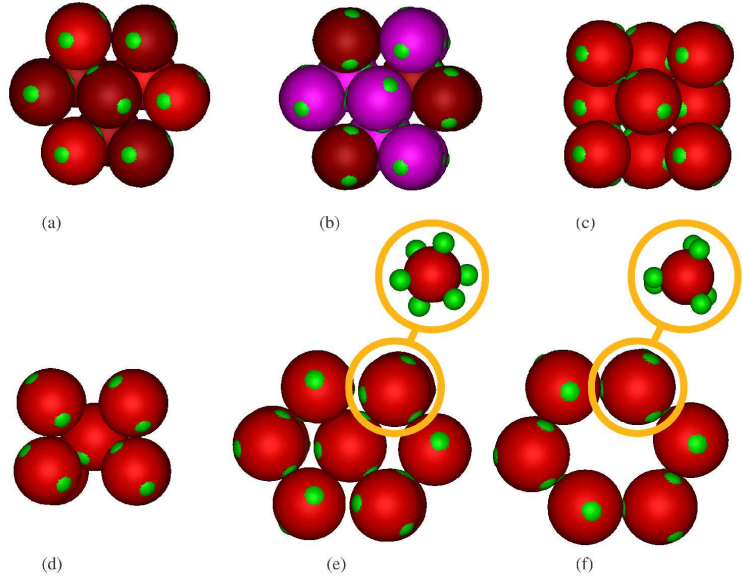}
\caption{(Colors online) Representative parts of the crystal structures identified for a particular realization of the Kern-Frenkel model with four patches per particle~\cite{KernModel03}.
 The colors of the particles represent the number of bonds per particle:
 light red = four bonds (i.e. fully bonded), magenta = three bonds, dark red = two bonds. 
 Sticky sites are colored in green.
  Packed structures:
  (a) and (b) HCP crystals with same average energy per particle and same average number density, 
  (c) fully bonded FCC crystal, 
  (d) fully bonded BCT crystal.
  Open structures:
  (e) part of a DC crystal and (f) part of a DH crystal. 
  The yellow circles highlight the different intra-layer bonding in the DC and the DH structures (in order to highlight the patch positions, a different 
  particle/patch size ratio is chosen; in each circle two particles, one on the top of the other, are reproduced): pairs of particles forming bonds between six-fold rings have 
  a staggered conformation in the DC crystal, and an eclipsed conformation in the DH crystal.
  Tab.~\ref{table:tab-kern} contains the  $e$- and $\rho$-values, as well as the $f$-values, of the respective candidate structures.}
\label{fig:kern}
\end{figure}

Representative parts of all the above mentioned structures are shown in Fig.~\ref{fig:kern}. 
Since in the model the number of bonds is well defined, the bond saturation is indicated in the figure via a color code.
For each structure, the corresponding values for the average energy per particle, $e$, and the average number density, $\rho$, 
are listed in Tab.~\ref{table:tab-kern}. The table also reports the frequency of appearance, $f$ (expressed in percentage), 
of the structures encountered in the MC simulations: 
out of a total of 160 parallel simulations at high/intermediate pressure values, each of them starting with different 
initial conditions, 90\% converged to one of the listed close-packed lattices. The corresponding values of $f$
for the open structures turns out to be significantly smaller: out of a total of 70 simulations at low pressure values,  
only 53\% converged to one of the two open configurations. This difference is due to the competition of the latter structures either with gel-like 
states or with hybrids between the DC and the DH lattices. It has been shown that, in large systems, hybrids of DC and DH structures
are predominant~\cite{Romano11}.

As shown in Tab.~\ref{table:tab-kern}, the DC and the DH crystals have the same $e$-  and $\rho$-values; 
however, in our simulation runs, we observe the DC structure with a slightly higher frequency than the DH lattice. 
Both diamond crystals are fully bonded structures built up of six-fold non-planar rings. The difference between the two four-coordinated 
particle arrangements can be clarified by inspecting the bonds between adjacent layers: 
as highlighted in Fig.~\ref{fig:kern} by the yellow circles, pairs of particles forming intra-layer bonds occur for the DC crystal in the staggered 
conformation and for the DH case in the eclipsed conformation~\cite{Romano11}.
At higher densities, the close-packed structure with the highest $f$-value is a fully bonded FCC crystal. 
This lattice can transform into another fully bonded, but more compact lattice, the BCT crystal, which can also 
be viewed as a face-centered crystal with a non-cubic unit cell. 
Finally, the best packed crystals found for this model are two structures of HCP type with the
same $e$- and $\rho$-values, but different bonding patterns; also the $f$-values of the two HCP lattices 
are considerably different. 

\begin{figure}[h]
\center
\includegraphics[width=12cm, clip=true]{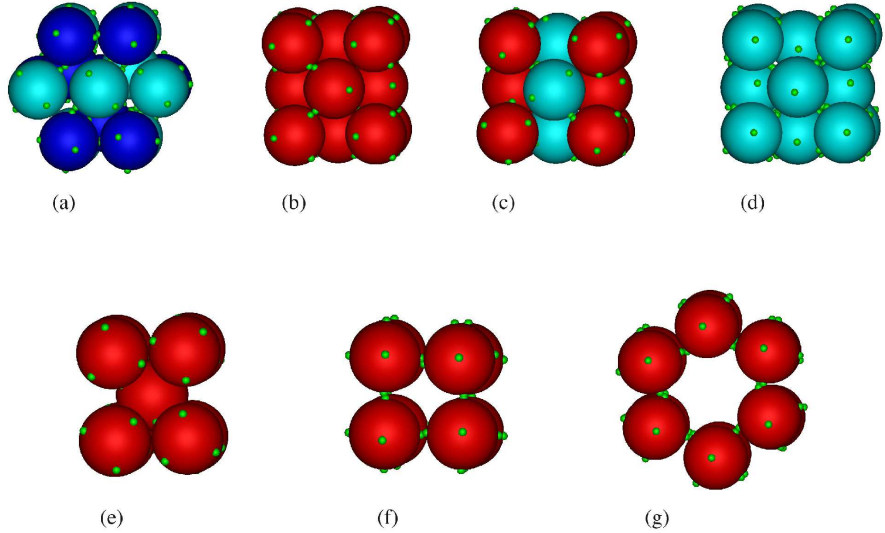}
\caption{(Colors online) Representative parts of crystal structures identified 
  for the sticky spots model with six patches per particle~\cite{Bianchi06}.
  The colors of the particles represent the number of bonds per particle: light red = six bonds (i.e. fully bonded), 
  blue = five bonds, turquoise = four bonds. Sticky sites are colored in green.
  Packed structures:
  (a) HCP structure,
  (b) fully bonded FCC crystal, 
  (c) and (d) partially bonded FCC crystal distinguished by different $e$- and $\rho$-values,
  and (e) fully bonded BCT.
  Fully bonded, open structures: 
  (f) SC crystal and
  (g) honeycomb layers (Hcl) (planes of six-fold planar rings).
  Tab.~\ref{table:tab-spots}  contains  the $e$- and $\rho$-values, as well as the $f$-values, of the respective candidate structures.}
\label{fig:sticky}
\end{figure}

Another patchy model of the orientational square-well type is the sticky spots model introduced in 
Ref.~\cite{Bianchi06}. Here, we consider particles decorated with six patches. 
To the best of our knowledge, the crystal phases of this model have not been investigated yet.
In Fig.~\ref{fig:sticky}, we show the unit cells of the candidate ordered structures. 
Since in the model the number of bonds is well defined, we make again use of a color code to indicate the bond saturation of each particle. 
In Tab.~\ref{table:tab-spots}, we report the corresponding $e$-, $\rho$-, and $f$-values for each structure.

As an open structure we consistently find the obvious Simple-Cubic (SC) crystal where all bonds are saturated. 
Additionally, we find (with a considerably lower $f$-value) another fully bonded structure, whose density is still smaller than that of
the SC lattice. Such an ordered structure is built up of parallel, connected planes, in which particles are arranged 
in a honeycomb (Hcl) geometry, i.e. as six-fold planar rings.
As candidate high pressure structures, we find HCP, FCC and  BCT crystals. 
The structures with the highest densities are a partially bonded HCP structure (see panel (a) of Fig.~\ref{fig:sticky})
and two FCC lattices (see panels (b) and (c) of Fig.~\ref{fig:sticky}), one of which is fully bonded,
while the other one is only partially bonded. The most frequently occurring high pressure lattice is a partially bonded FCC crystal 
(see panel (d) of Fig.~\ref{fig:sticky}), whose energy is higher than that of the other partially bonded FCC crystal and 
whose density is significantly smaller than that of the two FCC lattices mentioned above.
Finally, we identify the BCT crystal (see panel (e) of Fig.~\ref{fig:sticky}) as a fully bonded structure with a relatively
high density. 

We note that the percentage of simulations that converged to one of the close-packed structures listed in Tab.~\ref{table:tab-spots} adds up to 57\%,
while the corresponding total $f$-value for the open lattices is 50.6\%. The $f$-value for the 
low pressure simulations is comparable to the Kern-Frenkel case discussed above, indicating once more 
the competition between the open lattices and gel-like states. 
For the close-packed lattices, instead, the considerably lower success rate of the present model as compared to the Kern-Frenkel model
is related to the abundance of FCC structures with $e$ varying from the fully bonded case, i.e. $e=-3\epsilon$ (see panel (b) of Fig.~\ref{fig:sticky}) 
to $e=-2\epsilon$ (see panel (d) of Fig.~\ref{fig:sticky}). An example of a FCC structure with an intermediate $e$-value is shown 
in panel (c) of Fig.~\ref{fig:sticky}. 

\subsection{Orientational Lennard-Jones Models}
Next, we consider the orientational Lennard-Jones model with four and six patches per
particle. In our MC simulation, we succeed to identify all the structures reported 
in literature~\cite{Noya07,Noya2010}. 
For patchy particles carrying six patches on their surface, we find all and only the predicted phases, i.e. FCC, BCC and SC.
The corresponding unit cells are depicted in Ref.~\cite{Noya07}. 
In the four-patch case we identify the FCC and BCC phases reported in Ref.~\cite{Noya2010}. In addition, we observe the DC and the DH lattices.
The corresponding particle arrangements of the observed structures are similar to the ones displayed in panels (d)-(f) of Fig.~\ref{fig:kern}.
It has been shown~\cite{Noya2010} that the stability of the diamond structure sensitively depends on the
position of the potential minimum (i.e., on the optimal distance between two bonded particles) 
and consequently on the attractive interaction range.
For the chosen potential model, both diamond crystals are not thermodynamically stable~\cite{Noya2010}. 
This is an evidence of the need for full free-energy calculations to investigate the stability
of the candidate crystal structures found. 
            
\begin{figure}[h]
\centering
\includegraphics[width=11cm, clip=true]{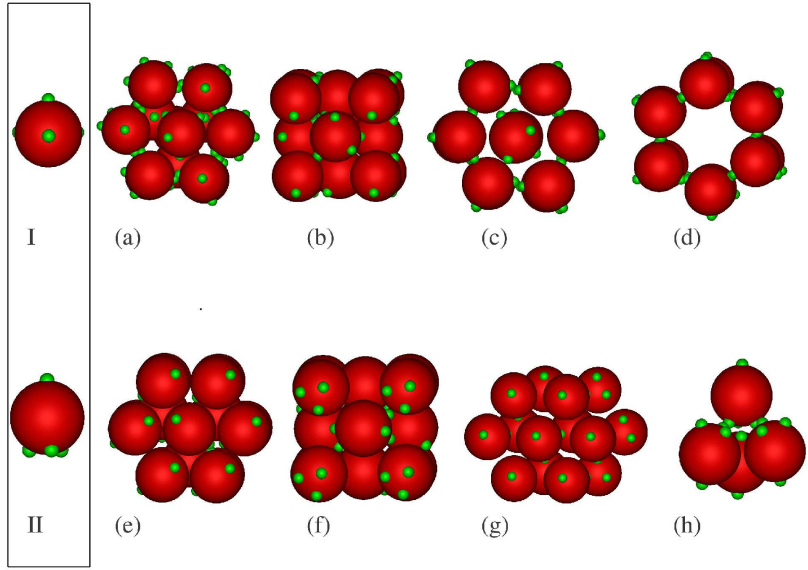}
\caption{(Colors online) Representative parts of crystal structures identified for the orientational
  Lennard-Jones model with four patches per particle introduced in Ref.~\cite{Doppelbauer2011}.
  The selected patch geometries are displayed in the rectangular box on the left hand side: 
  case ${\rm I}$ corresponds to the compressed tetrahedron with $g=90^{\circ}$, and
  case ${\rm II}$ corresponds to the elongated tetrahedron with $g=150^{\circ}$.
  Representative parts of the observed structures for case  ${\rm I}$ : 
  (a) HCP crystal,
  (b) FCC crystal,
  (c) hexagonal double layers (HxDl), and
  (d) honeycomb double layers (HcDl).
  Representative parts of  the observed structures for case  ${\rm II}$: 
  (e) HCP crystal,
  (f) FCC crystal, 
  (g) hexagonal staggered double layers (HxSDl),
  and (h) tetrahedrally arranged lattice (Ts).
  Tab.~\ref{table:tab-ga} contains the $e$- and $\rho$-values, as well as the $f$-values, of the respective candidate structures.}
\label{fig:ga}
\end{figure}

We also consider a related orientational Lennard-Jones model with four patches arranged in different tetrahedral 
geometries on the particle surface~\cite{Doppelbauer2011}, specified by the geometrical parameter $g$. 
Here, we show results for two extreme patch arrangements, 
i.e. $g=90^{\circ}$ (compressed tetrahedron) and $g=150^{\circ}$ (elongated tetrahedron) (see left column of Fig.~\ref{fig:ga} for a schematic representation). 
We compare candidate structures proposed by the variable box shape MC simulation technique and lattices suggested by the evolutionary algorithm approach. 
The candidate structures found via MC simulations are shown in Fig.~\ref{fig:ga}; the corresponding $e$-, $\rho$-, and $f$-values
of the structures are listed in Tab.~\ref{table:tab-ga}.

In the case of compressed tetrahedrons ($g=90^{\circ}$),  we observe, at low pressure values, the formation of either honeycomb double layers (HcDl) 
or hexagonal double layers (HxDl). The HcDl structure is characterized by a fully bonded, planar honeycomb lattice, where the bonds within 
the six-fold rings are formed by the patches located on the equatorial plane of the particles. The remaining patches of each of the six particles 
forming a ring are all oriented in the same direction, providing the intra-layer link: two oppositely oriented layers are connected via these intra-layer linkers,
forming hereby double layers. 
Between the double layer, there is no attractive interaction, thus they can either be far from each other or 
they can be almost in contact (the density value reported in Tab.~\ref{table:tab-ga} 
refers to the latter case). The HxDl structure can be viewed as a HcDl structure with an additional particle located in the center of the six-fold honeycomb ring. 
In this case, double layers do interact with each other since the central particles provide links between the double layers, leading thereby to a higher $\rho$-value
as compared to the preceding case. 
The $e$-value for the HxDl lattice is considerably higher than that of the HcDl, since, in this particle arrangement, the six-fold rings are slightly distorted in order 
to appropriately accommodate the seventh particle in their center.
At high/intermediate pressure values, the most frequently encountered structure is the FCC crystal. A slightly more compact structure that is observed is a HCP crystal 
with higher $e$- and $\rho$- values.
We note that, for this particular model, the $f$-values are throughout higher than those reported for the square-well patchy models. 
To be more specific, the percentage of simulations that converged to any of the packed structures at high/intermediate pressure values amounts to 97\%, 
while the percentage of low pressure runs that converged to any of the open structures at low pressure values is 76\%. 

Among the structures obtained from the MC simulations, only two are identified as global minima of the Gibbs free energy at zero temperature
by the evolutionary algorithm method: the HcDl, at low pressure values, and the FCC crystal, at high pressure values. 
The additional crystals obtained via the MC simulations, i.e. the HxDl and the HCP structures, are identified as low-lying local minima on the Gibbs free energy landscape by the evolutionary algorithm. In Tab.~\ref{table:tab-ga}, we report the corresponding $e$- and $\rho$-values of the structures identified by the evolutionary algorithm together with the information about the location of the corresponding minimum on the Gibbs free energy landscape at zero temperature.
In the intermediate range of pressure, the evolutionary algorithm identifies two additional global minima corresponding to two different kinds of hexagonal double layers; they are not
found via the MC simulation technique. In Tab.~\ref{table:tab-ga}, we refer to them as HxDl-I and HxDl-II. For the visualization of such hexagonal double layers, see Ref.~\cite{Doppelbauer2011}.

In the case of elongated tetrahedrons ($g=150^{\circ}$), we observe, at low pressure values, the formation of a tetrahedrally arranged 
structure (Ts) (see panel (h) of Fig.~\ref{fig:ga}) as well as the formation of a slightly more compact lattice built up of staggered and connected double layers with 
a hexagonal structure (HxSDl) (see panel (g) of Fig.~\ref{fig:ga}), which is characterized by a slightly lower energy than the Ts configuration.  
The bonding patterns of the two lattices are distinctively different. In the Ts lattice each patch on a particle is strongly interacting with a patch on a neighboring particle: 
in the part of the Ts structure shown in Fig.~\ref{fig:ga}, the upper particle and one of the lower particles are oriented with the polar patch pointing upwards, 
while the other two particles at the bottom are oriented downwards. 
In the HxSDl lattice, instead, pairs of oppositely oriented layers are bound to each other via the polar patches of each of the seven particles forming the hexagonal tiling;
however, among the remaining patches, only two of them (per particle) are strongly interacting with the corresponding patch on the other, oppositely oriented layer.
At high and intermediate pressure values, we identify either FCC or HCP crystals: the FCC structure is
the most probable one; on the other hand, the HCP crystal maximizes the density but has a significantly lower $e$-value in comparison to
the FCC lattice. Once more, we note that, for this model, the frequencies of occurrence sum up to values significantly higher than those reported for the square-well type patchy models:
99\% for the packed structures and 75\% for the open lattices.
 
As we compare the results from the two different methods, we note that all the global minima identified by the evolutionary algorithm at zero temperature, i.e.  the HCP and the HxSDl structures, are also obtained from the MC simulations at low (but finite) temperature.  In Tab.~\ref{table:tab-ga}, we report the corresponding $e$- and $\rho$-values of the two structures identified by the evolutionary algorithm. None of the additional crystals obtained from the MC simulations, neither the FCC nor the Ts lattices, are identified as global minima of the Gibbs free energy at zero temperature. Nonetheless,  both structures, at high and low pressure respectively, are among the best configurations
found by the optimization technique: they are both identified as low-lying local minimum of the Gibbs free energy, 
in their corresponding range of pressure. 

\begin{figure}[h]
\centering
\includegraphics[width=14cm, clip=true]{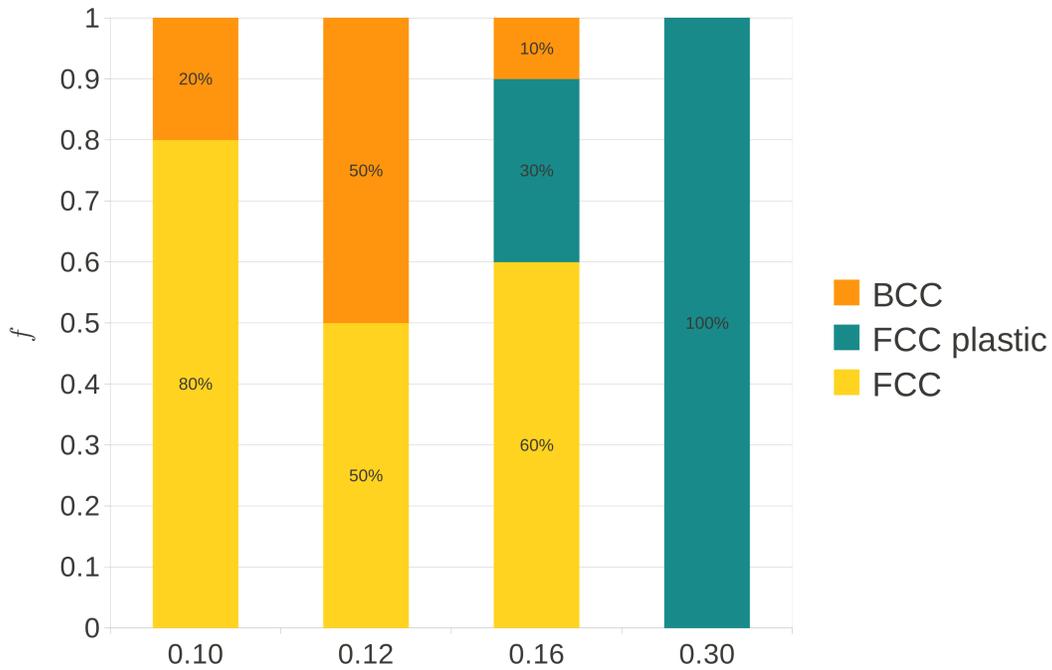}
\caption{Overview of the packed crystal structures for the orientational Lennard-Jones model with four patches per particle arranged in a regular tetrahedral geometry ($g=109.47^{\circ}$). The histogram reports the $f$-values (normalized to unity) of each identified structure at different temperatures (in units of $\epsilon/k_B$) indicated at the bottom of each bar.}
\label{fig:tetra}
\end{figure}

To conclude, we also consider the regular tetrahedral case corresponding to $g=109.47^{\circ}$. The phase diagram of such a system is reported in Ref.~\cite{Doppelbauer2011}; it shows the presence of three different FCC-like phases: a low temperature face-centered non-cubic structure, an FCC phase, and a plastic FCC crystal (FCC$_p$). The EA approach only identifies the first one as the global enthalpy minimum at zero temperature; however, the FCC lattice is identified as a local minimum of the enthalpy landscape~\cite{Doppelbauer-LMC2011}. In contrast, the MC approach is able to identify the latter two FCC-like phases. In Fig.~\ref{fig:tetra}, we compare the frequencies of appearance of the packed structures found via the variable box shape simulation technique over a large temperature range. Among the packed structures, we also identify a BCC like phase, which has a wide region of stability in the phase diagram~\cite{Doppelbauer2011}. Fig.~\ref{fig:tetra} shows that the variable box shape simulation technique is able to properly take into account the effect of temperature; indeed at the highest temperatures investigated, FCC$_p$ is observed with increasing frequency. On the other hand, with the MC approach, we do not observe the non-cubic structure, whose region of stability is confined to very low temperatures~\cite{Doppelbauer2011}.

\section{Conclusions}\label{sec:conclusions}

In this paper we have employed the MC NPT variable box shape simulations to predict candidate structures for several patchy particle models proposed in the literature.
We have determined the structures for two patchy hard-sphere models, i.e. the Kern-Frenkel model with four patches and the ``sticky spots'' model with six patches.
For the Kern-Frenkel model, we find all the stable phases as previously predicted in Ref.~\cite{Romano10,Romano11}, thereby giving confidence in the MC method. Moreover, we find a BCT and two HCP phases which stability should be determined by free energy calculations. For the sticky spots model, we have successfully predicted several candidate structures. To the best of our knowledge, the crystal phases of this system have not been studied before.

In addition, we have compared crystal structures predicted by MC NPT
variable box shape simulations with results obtained via an
evolutionary algorithm approach for various patchy particle systems interacting via continuous pair potentials.
From our findings, it appears that neither method is significantly
better than the other, and that the most appropriate method for a
given system depends on the characteristic  features of the
problem.  
Concerning the relative efficiency in finding the solid structures, approximately 90\% of the EA runs converge to the same minimum of the enthalpy landscape, irrespective whether it corresponds to an open or a packed configuration. On the other hand, MC runs show two different percentage values in the two cases: almost 100\% in the high/intermediate pressure regime and around 75\% in the low pressure regime. Moreover, the computational costs of the two approaches differ by less than one order of magnitude on a real time scale. In the following, we briefly highlight a few of our considerations on the comparison between the two numerical approaches:
\begin{itemize}
\item[(i)] For low temperatures ($T\sim 0$), the two techniques
  produce virtually equivalent crystal structures, when a selection of
  the lowest-lying local minima on the enthalpy landscape identified
  by the evolutionary algorithm is taken into
  account:  
  structures listed among the energetically most favorable ones according to the evolutionary algorithm technique,
  despite not being a global minimum at zero temperature, can be possibly thermodynamically stable at finite temperatures~\cite{Doppelbauer2011}.
  This encouraging fact is a hint
  on the reliability of both methods. 
  Structures that are identified as $T=0$ stable phases 
  can be then considered as good candidates at finite (even though low) temperature.
\item[(ii)] Further, the MC method can predict candidate crystal
  phases at finite temperature, unlike the evolutionary algorithm,
  which is bound to $T=0$  imposed by conceptual and computational
    limitations. Indeed, on increasing the temperature beyond the chosen range ($T \geq 0.2$), we identify FCC plastic crystal phases for all the discussed models. Moreover for the regular tetrahedral Lennard-Jones patchy system we are able to identify a FCC crystal structure which is stable only at finite temperature and cannot be found via the EA approach.
  On the other hand,  at the temperatures relevant for patchy systems the
  simulations are more likely to get kinetically trapped in gel-like
  states or non-competitive local minimum configurations, which have
  to be discarded. This problem is much easier to handle within the
  evolutionary algorithm by using suitably designed ``population
  control'' operations~\cite{Hartke99,Doppelbauer2011}.
\item[(iii)] For systems with discontinuous interaction potentials,
  the MC method has the advantage that it can be applied directly,
  while for use with an evolutionary algorithm either suitable approximations (by smoothening the potential) or cumbersome methodological implementations~\cite{Gernot2009_bis} are needed.  
\end{itemize}

We stress again, that the thermodynamic stability of the
crystal structures predicted by both methods is not guaranteed and has
to be checked by full free-energy
calculations~\cite{Marechal-2010,Marechal-2011,VegaSAN08,Doppelbauer2011}.

As the two methods covered here have both advantages as well as
shortcomings, we list potential improvements in the following. A
possibility to overcome kinetic trapping in gel-like states for
variable box-shape simulations is to combine this method with moves
that correspond to larger leaps in configuration space, comparable to
mating and mutation operations of an evolutionary algorithm; this
would move the method further away from being a thermodynamic approach
into the direction of an optimization technique. The evolutionary
algorithm, on the other hand, could be augmented with especially
effective mutation steps, that are based on short MC runs, as
suggested in Ref.~\cite{Hartke99}. Ultimately, even a hybrid approach,
incorporating the advantages of both methods, is conceivable.  Another
desirable amendment to EAs would be using free energy calculations
based on lattice dynamics\cite{TaylorBAB97} in order to estimate the
competitiveness of candidate structures at {\it finite} temperature
already during the run of the algorithm; it has to be noted though,
that such an approach demands the interaction potential to meet even
stronger criteria (continuous second derivatives) and is conceptually
rather involved and computationally expensive.

\section{Acknowledgements}

The authors thank Michiel Hermes for technical support. EB wishes to
thank the Austrian Research Fund (FWF) for support through a
Lise-Meitner Fellowship under project number M1170-N16 as well as
the Deutsche Forschungsgemeinschaft SFB-TR6 program for additional support. 
Further financial support by the FWF under the project numbers W004 and P23910-N16
is gratefully acknowledged.


%

\clearpage

\begin{table}[htdp]
\begin{center}
\begin{tabular}{c|c|c|c}
\hline 
\hspace{0.1cm} structure \hspace{0.5cm}&\hspace{0.1cm}  $e$ \hspace{0.5cm}&
\hspace{0.1cm} $\rho$    \hspace{0.5cm}&\hspace{0.1cm}  $f$ \\
\hline 
\hline
HCP (a)  & -4/3  & 1.37 &  8\% \\
HCP (b)  & -4/3  & 1.37 &  4\%   \\
FCC (c)  & -2    & 1.18 &  53\%  \\
BCT (d)  & -2    & 1.25 &  26\% \\
\hline
DC  (e)  & -2    & 0.6 &  30\% \\
DH  (f)  & -2    & 0.6 &  23\% \\    
\hline 
\end{tabular}
\end{center}
\caption{Overview of the crystal structures shown in Fig.~\ref{fig:kern} for the four patch Kern-Frenkel model.
For each structure, the corresponding values of the average 
energy per particle $e$ (in units of $\epsilon$), the average density $\rho$ (in units of $\sigma^{-3}$), and the frequency of
appearance $f$ (\%) are reported.
The upper part of the table contains the four structures found for high/intermediate pressure values (over a total of 160 different simulations 
started in parallel with different initial conditions); the part of the table below the horizontal line contains the two open structures found at low pressure
values (over a total of 70 different simulations). The densities of the DC and the DH crystals are reported with a one-digit precision (vs a two-digits precision in the 
case of close-packed crystals) since, for a square well attraction of range $\delta=0.24\sigma$, the density of an open 
bonded structure can vary up to 10\% without causing an additional cost in energy.}
\label{table:tab-kern}
\end{table}

\clearpage

\begin{table}[htdp]
\begin{center}
\begin{tabular}{c|c|c|c}
\hline 
\hspace{0.1cm} structure \hspace{0.5cm}&\hspace{0.1cm}  $e$ \hspace{0.5cm}&
\hspace{0.1cm} $\rho$    \hspace{0.5cm}&\hspace{0.1cm}  $f$ \\
\hline 
\hline
HCP (a)  & -2.25 &  1.37 &  1\% \\
FCC (b)  & -3    &  1.37 &  3\% \\
FCC (c)  & -2.5  &  1.37 &  7\% \\
FCC (d)  & -2    &  1.18 &  38\% \\
BCT (e)  & -3    &  1.33 &  8\% \\
\hline
SC  (f)  & -3    &  0.87 &  44\% \\
Hcl (g)  & -3    &  0.80 &  7\% \\    
\hline 
\end{tabular}
\end{center}
\caption{Overview of the crystal structures shown in Fig.~\ref{fig:sticky} for the sticky spots model with six patches per particle.
For each structure, the corresponding values of the average 
energy per particle $e$ (in units of $\epsilon$), the average density $\rho$ (in units of $\sigma^{-3}$), and the frequency of
appearance $f$ (\%) are reported. 
The upper part of the table contains the five structures found for high/intermediate pressure values (over a total of 315 simulations); 
the lower part of the table below the horizontal line contains the two open structures found at low pressure values (over a total of 225 simulations). 
The densities of the open structures are reported with a two-digits precision (as in the case of close-packed crystals), 
since the change in density -- without additional energy costs -- of an open bonded structure 
can amount to 6\% at maximum in the case of a square well attraction of range $\delta=0.119\sigma$.}
\label{table:tab-spots}
\end{table}

\clearpage

\begin{table}[htdp]
\begin{center}
\begin{tabular}{c|c|c|c|c|c|c|c|c}
\hline 
Geometry  &\hspace{0.1cm}  structure \hspace{0.5cm}&\multicolumn{3}{c|}{MC} &\multicolumn{4}{c}{EA}\\ 
\hline
          &\hspace{0.1cm}            \hspace{0.5cm}&\hspace{0.1cm}  $e$ \hspace{0.5cm}&\hspace{0.1cm} $\rho$ \hspace{0.5cm}&\hspace{0.1cm}  $f$ 
                                     \hspace{0.5cm}&\hspace{0.1cm}  $e$ \hspace{0.5cm}&\hspace{0.1cm} $\rho$ \hspace{0.5cm}&\hspace{0.1cm}  $P$ \hspace{0.5cm}&\hspace{0.1cm}  minimum \\
\hline 
\hline
${\rm I}$  & HCP  (a)      & -0.83 & 1.23  &  17\% & -1.19 & 1.34 & 9.1 & LM$_1$ \\
${\rm I}$  & FCC  (b)      & -1.03 & 1.18  &  80\% & -1.23 & 1.32 & 9.1 & GM     \\
${\rm I}$  & HxDl-II       &   &  &                & -1.18 & 1.35 & 6.0 & GM     \\
\hline
${\rm I}$  & HxDl (c)      & -1.17  & 0.91 &  24\% & -1.68 & 0.98 & 1.1 & LM$_5$ \\
${\rm I}$  & HxDl-I        &   &  &                & -1.84 & 1.04 & 1.1 & GM     \\
${\rm I}$  & HcDl  (d)      & -1.50  & 0.57 & 50\% & -2.00 & 0.65 & 0.1 & GM     \\
\hline
\hline
${\rm II}$ & HCP  (e)  & -1.13  & 1.30 &  33\% & -1.69 & 1.33 & 6.0 & GM     \\
${\rm II}$ & FCC  (f)  & -0.88  & 1.18 &  66\% & -1.66 & 1.32 & 6.0 & LM$_1$ \\
\hline
${\rm II}$ & HxSDl (g)  & -1.63  & 1.02 & 45\% & -2.00 & 1.11 & 0.1 & GM \\
${\rm II}$ & Ts    (h)  & -1.54  & 0.88 & 30\% & -2.00 & 1.03 & 0.1 & LM$_2$\\
\hline 
\end{tabular}
\end{center}
\caption{
Overview of the crystal structures for the orientational Lennard-Jones model with four patches per particle. 
Structures identified by both the Monte Carlo (MC) and the evolutionary algorithm (EA) approach are listed according to the labels of Fig.~\ref{fig:ga}; for the structures that are obtained only from the EA method we refer to Ref.~\cite{Doppelbauer2011}.
The first column of the table specifies the patch geometry: 
case ${\rm I}$ corresponds to $g=90^{\circ}$, while case ${\rm II}$ corresponds to $g=150^{\circ}$. 
The values of $e$ (in units of $\epsilon$), $\rho$ (in units of $\sigma^{-3}$) and $f$ for the candidate structures obtained via the MC simulations
are shown in the third, fourth and fifth columns, respectively.
The sixth and the seventh columns report the $e$- and $\rho$-values of the structures as identified by the EA approach; these values are obtained at the pressure reported in the 8th column. When a given structure is (locally) optimized for a different pressure value, $e$ and $\rho$ can change by about 5\%. The ninth column gives information about the location of the corresponding minimum on the Gibbs free energy landscape at $T=0$: GM corresponds to the global minimum and LM$_{n}$ corresponds to the $n_{th}$ local minimum after the global one.
For both patch geometries, the upper part of the table includes structures at high/intermediate pressure values (41 MC simulations for case ${\rm I}$, 15 MC simulations for case ${\rm II}$) and the lower part contains open structures at low pressures (50 MC simulations in case ${\rm I}$, 20 MC simulations in case ${\rm II}$).
}
\label{table:tab-ga}
\end{table}

\clearpage

\end{document}